# On Non-Consensus Motions of Dynamical Linear Multi-Agent Systems

Cai, Ning[1, 2, 4]    Deng, Chun-Lin[1, 3]    Wu, Qiu-Xuan[4]

[1]School of Electrical Engineering, Northwest University for Nationalities, Lanzhou, China
[2]Key Laboratory of National Language Intelligent Processing, Gansu Province, China
[3]School of Automation, Huazhong University of Science and Technology, Wuhan, China
[4]School of Automation, Hangzhou Dianzi University, Hangzhou, China

**Abstract:** The non-consensus problems of high order linear time-invariant dynamical homogeneous multi-agent systems are concerned. Based on the conditions of consensus achievement, the mechanisms that lead to non-consensus motions are analyzed. Besides, a comprehensive classification for diverse types of non-consensus phases in accordance to the different conditions is conducted, which is jointly depending on the self-dynamics of agents, the interactive protocol and the graph topology. A series of numerical examples are demonstrated to illustrate the theoretical analysis.

## 1. Introduction

The dynamical multi-agent systems, as a type of large-scale complex system, are composed of numerous autonomous or semi-autonomous subsystems, being relatively equipotent and connected through the information interacting network. During recent decades, the consensus problem of dynamical multi-agent systems has attracted extensive attention in control theory. The basic idea of consensus is that a group of agents achieve an agreement over some variables of interest by local interactions.

Olfati-Saber *et al.* introduced the term "consensus" into the control theory firstly [1]. Ren and Beard [2] relaxed the conditions in [1], and pointed out the fact that the communication topology has a spanning tree is critical for a multi-agent system to achieve consensus. A set-valued Lyapunov function method was developed by Moreau [3]. Until 2007, the majority of studies on consensus problem had dealt with first-order models. Since 2007, consensus problems for high-order multi-agent systems have been addressed. For instance, Xiao *et al.* [4] proposed a criterion based

---

on the structure of certain high-dimensional matrices. Wang *et al.* [5] attempted to determine whether an appropriate linear high-order consensus protocol exists under a given undirected graph topology. Cai *et al.* proved necessary and sufficient conditions for both swarm stability and consensus of high-order LTI (Linear Time-Invariant) normal systems [6], nonlinear systems [7], and singular systems [8], respectively. Li *et al.* studied the robust stability problem of linear multi-agent systems with observer type interactive protocols [9], and later developed an approach [10] to realize consensus using merely the local information, without knowing the global structure of the network topology. Xi *et al.* devised a technique based on oblique decomposition of state space [11], and recently addressed the guaranteed cost control for consensus [12], which can be regarded as a suboptimal control problem for interconnected systems. Other relevant works are included in [13-18].

The majority of past studies mainly focused on discovering the approaches and conditions to achieve consensus for different multi-agent systems under various situations. The reasons why so much attention has been paid to the problem of achieving consensus are that: 1) The consensus is a specific and relatively simple case of the stability of multi-agent systems, also being referred to as asymptotic swarm stability; meanwhile, a conventional viewpoint in control theory is that stability is prerequisite for systems to operate normally; 2) Some other more complicated control problems can be transformed into consensus under certain conditions, e.g. the formation control [19].

However, it is noteworthy that consensus is only a particular example of the stability of multi-agent systems. Actually in many practical applications, mere consensus is sometimes restrictive. For example, in most cases, the formation control [19], flocking control [20], containment control [21] and some other control problems are free of consensus and cannot be transformed into equivalent consensus-based problems. Evidently, compared with consensus, non-consensus is more common in various practical applications, containing broader generality. In addition, the consensus could be even harmful to the overall stability of some real systems. For instance, the state of system will oscillate seriously when the consensus of certain variables occurs in many economic systems [22], and thus consensus should instead be avoided deliberately. Hence, it is time to extend our perspective on consensus to the study of diverse non-consensus cases, which are much more complicated and challenging.

A few studies on the conditions and methods for some non-consensus problems have been conducted so far, e.g. group clustering. Yu *et al.* [23] concerned the group consensus in multi-agent systems with switching topologies and communication delays. Hu *et al.* [24] studied a group consensus problem with discontinuous information transmissions among different groups of dynamical agents. Su *et al.* investigated the pinning control for cluster synchronization of undirected complex dynamical networks using a decentralized adaptive strategy [25] and the cluster synchronization of coupled harmonic oscillators with multiple leaders in an undirected fixed network [26].

The study on the non-consensus problem in depth not only can lay a solid theoretical foundation for many practical applications, but can also further facilitate our deeper understanding on the consensus problems. However, no papers exist hitherto which aim to systematically study the motions being free of consensus and reveal the mechanism of non-consensus. In this paper, the non-consensus problem of LTI multi-agent systems will be addressed. Unlike many existing investigates on the non-consensus problem which just concerned certain particular motion paradigm, e.g. the group clustering, a comprehensive classification of different types of non-consensus phases along with the analysis for corresponding conditions will be provided. In fact, one might discover that the mechanism of non-consensus, which mainly depends on the graph topology, the agent dynamics, and the interactions of agents, is less complicated than imagination, with different non-consensus motions of LTI multi-agent systems being classified into two categories roughly: group clustering and mutual repulsion between any two agents. Concretely speaking, the features of non-consensus dynamics can be assigned into three primary paradigms in accordance with the different conditions around consensus achievement.

The main contribution of this paper is comprehensive classification and analysis of the conditions for various types of non-consensus motions of LTI multi-agent systems, illustrated by several typical numerical instances. The discussions here can deepen our understanding of the mechanisms about the dynamics of multi-agent systems, both consensus and non-consensus.

The rest of this paper is organized as follows. In Section 2, the mathematical model of homogeneous high-order LTI multi-agent systems and relevant fundamentals are introduced. Section 3 provides a comprehensive classification for the paradigms of non-consensus motions, with both empirical and theoretical analysis.

Finally, Section 4 concludes this paper.

## 2. Model Formulation and Preliminaries

In this paper, the mathematical model of homogeneous high-order LTI multi-agent systems takes the following form:

$$\begin{cases} \dot{x}_1 = Ax_1 + F\sum_{j=1}^{N} w_{1j}(x_j - x_1) \\ \dot{x}_2 = Ax_2 + F\sum_{j=1}^{N} w_{2j}(x_j - x_2) \\ \vdots \\ \dot{x}_N = Ax_N + F\sum_{j=1}^{N} w_{Nj}(x_j - x_N) \end{cases} \quad (1)$$

where $x_i \in R^d$ ($i \in \{1, 2, \cdots, N\}$) denotes the state vector of agent $i$; $A, F \in R^{d \times d}$ represent the autonomous dynamics of each agent and interactive dynamics among agents, respectively; $w_{ij} \in R^+$ denotes the arc weight between agent $i$ and $j$ in the graph of system $G$. The adjacency matrix of $G$ is

$$G:W = \begin{bmatrix} w_{11} & w_{12} & \cdots & w_{1N} \\ w_{21} & w_{22} & \cdots & w_{2N} \\ \vdots & \vdots & \vdots & \vdots \\ w_{N1} & w_{N2} & \cdots & w_{NN} \end{bmatrix}$$

This matrix is symmetric if and only if the graph is undirected, otherwise it is asymmetric. Weighted adjacency matrix is more general than the case with binary 0-1 element values, with the weight $w_{ij}$ being regarded as the strength of information link between agents $i$ and $j$.

It is worth mentioning that many practical engineering systems can be described with the model, e.g. the multi-agent supporting systems (MASS) [27].

It has been a common knowledge that consensus of a multi-agent system implies joint convergence, which is formulated by the following definition.

*Definition 1:* For system (1), if

$$\lim_{t \to \infty} \|x_i(t) - x_j(t)\| = 0$$

then agents $i$ & $j$ achieve an *agreement*. For a given vertex set $V_k \subset V$, if $\forall i, j \in V_k$, agents $i$ & $j$ achieve agreement, then the multi-agent system (1)

achieves a *consensus* in $V_k$. If consensus is achieved in $V_1$, $V_2$, …, $V_\alpha$, respectively, with

$$V_1 \cup V_2 \cup \cdots \cup V_\alpha = V$$

then the overall system achieves a *group consensus*.

On the basis of the above mathematical model of high-order LTI multi-agent systems and the definition of consensus, many studies on the conditions and approaches for achieving consensus have been conducted. The well-known consensus criterion of high-order LTI multi-agent systems is a fundamental for studying the non-consensus problem of LTI multi-agent systems in this paper, which is illustrated as follows:

*Lemma 1* [6]: For multi-agent system (1) with $\lambda_1, \lambda_2, \cdots, \lambda_N \in C$ as the eigenvalues of the Laplacian matrix $L(G)$, if $A$ is not Hurwitz, the system achieves consensus iff both conditions 1 and 2 below are true:

1. The graph topology $G$ includes a spanning tree;
2. All the matrices $A - \lambda_i F$ ($\lambda_i \neq 0$) are Hurwitz.

If $A$ is Hurwitz, then the system achieves consensus iff condition 2 is true.

## 3. Classification and Numerical Simulation on Non-Consensus Motions

According to the criterion on checking consensus of high-order LTI multi-agent systems, various conditions of non-consensus motions can be classified into three primary classes. The main theme of the current section is to elaborate the different classes by theoretical analyses and simulations.

### 3.1. Class 1

The situation of Class 1 of non-consensus motions of high-order LTI multi-agent systems can be summarized as:

1. Some of the elements in

$$\{A - \lambda_i F \mid i = 1, 2, ..., N; \lambda_i \neq 0\}$$

   are not Hurwitz, where $\lambda_i$ ($i = 1, 2, ..., N; \lambda_i \neq 0$) are the nonzero eigenvalues of the Laplacian matrix $L(G)$;

2. At least a spanning tree is included in the graph of system.

*Example 1*: Consider an LTI multi-agent system with the graph containing a spanning tree, which is illustrated in Fig. 1.

>>put Fig. 1 about here<<

Fig.1. Graph of system. Default edge weight is 1.

The adjacency matrix is

$$W = \begin{bmatrix} 1 & 1 & 1 & 1 & 0 & 0 \\ 1 & 1 & 1 & 0 & 0 & 0 \\ 0 & 1 & 0 & 0 & 0 & 1 \\ 0 & 1 & 1 & 0 & 1 & 1 \\ 1 & 0 & 0 & 0 & 0 & 0 \\ 0 & 0 & 0 & 1 & 1 & 1 \end{bmatrix}$$

and the Laplacian matrix is

$$L = \begin{bmatrix} 3 & -1 & -1 & -1 & 0 & 0 \\ -1 & 2 & -1 & 0 & 0 & 0 \\ 0 & -1 & 2 & 0 & 0 & -1 \\ 0 & -1 & -1 & 4 & -1 & -1 \\ -1 & 0 & 0 & 0 & 1 & 0 \\ 0 & 0 & 0 & -1 & -1 & 2 \end{bmatrix}$$

with the eigenvalues $\{3.96+0.56i, 3.96-0.56i, 0, 1.53+0.51i, 1.53-0.51i, 3\}$.

Suppose that the agents are LTI second-order asymptotically stable systems with

$$A = \begin{bmatrix} -1 & 1 \\ 0 & -2 \end{bmatrix} \text{ and } F = \begin{bmatrix} -0.5 & 0.5 \\ -0.5 & -0.5 \end{bmatrix}$$

For the above LTI multi-agent system, the eigenvalues of state matrix $A$ are $\{-1, -2\}$, and the eigenvalues of $A - \lambda_i F$ $(i = 1, 2, \ldots, 6)$ are

$$\{0.80 - 1.03i, 0.17 + 1.60i\}$$
$$\{0.80 + 1.03i, 0.17 - 1.60i\}$$
$$\{-1.00, -2.00\} \quad (\lambda_3 = 0)$$
$$\{-0.02 + 0.16i, -1.44 + 0.35i\}$$
$$\{-0.02 - 0.16i, -1.44 - 0.35i\}$$
$$\{\pm 0.71i\}$$

respectively. Evidently, some of the elements in $\{A - \lambda_i F \mid i = 1, 2, \ldots, N; \lambda_i \neq 0\}$ are not Hurwitz. The state trajectories of Example 1 are shown in Fig. 2.

>>put Fig. 2 about here<<

Fig. 2. State trajectories of Example 1 with $t \in [0, 7]$. Thick dots denote starting positions.

In Fig. 2, one can see that each agent is mutually repulsive to others in the phase diagram such that the relative states among agents will gradually enlarge as time elapses.

*Example 2*: Consider an LTI multi-agent system also with the graph shown in Fig. 1.

Suppose that the agents are LTI second-order unstable systems with

$$A = \begin{bmatrix} 1 & 1 \\ -2 & 0 \end{bmatrix} \text{ and } F = \begin{bmatrix} -0.65 & -1.65 \\ 0.07 & 0.40 \end{bmatrix}$$

For the above LTI multi-agent system, the eigenvalues of state matrix of each agent $A$ are $\{0.50 \pm 1.32i\}$, and the eigenvalues of $A - \lambda_i F$ $(i = 1, 2, \ldots, 6)$ are

$$\{0.86 + 3.32i, \ 1.13 - 3.18i\}$$
$$\{0.86 - 3.32i, \ 1.13 + 3.18i\}$$
$$\{0.50 \pm 1.32i\} \quad (\lambda_3 = 0)$$
$$\{0.44 + 2.48i, \ 0.94 - 2.35i\}$$
$$\{0.44 - 2.48i, \ 0.94 + 2.35i\}$$
$$\{0.88 \pm 2.97i\}$$

respectively. Evidently, all of the elements in $\{A - \lambda_i F \mid i = 1, 2, \ldots, N; \ \lambda_i \neq 0\}$ are not Hurwitz. The state trajectories of Example 2 are shown in Fig. 3.

>>put Fig. 3 about here<<

Fig. 3. State trajectories of Example 2 with $t \in [0, 6.6]$. Thick dots denote starting positions.

In Fig. 3, one can see that each agent is mutually repulsive to others in the phase diagram such that the relative states between agents will gradually enlarge as time elapses.

Obviously, the state trajectories of systems in both Example 1 and Example 2 are divergent, i.e. all agents in system are mutually repulsive.

In fact, the feature of motions for Class 1 of LTI multi-agent systems is complicated and indefinite, sensitively depending on the specific values of matrices $A$, $F$ and $L$.

It is a notable feature for condition in Class 1 that there may appear quasi-clustering phenomenon. This subsection can be concluded by a necessary and sufficient condition for checking whether or not a clustering phenomenon will occur.

*Lemma 2 [28]:* Consider the dynamical system (1). Suppose that the spectrum of Laplacian matrix of the directed graph with spanning tree is
$$\{\lambda_1 = 0, \lambda_2, ..., \lambda_N\}$$
with the series of matrices
$$A, A - \lambda_2 F, ..., A - \lambda_{\alpha-1} F$$
being not Hurwitz, and
$$A - \lambda_\alpha F, A - \lambda_{\alpha+1} F, ..., A - \lambda_N F$$
being Hurwitz. The pair of agents $i$ and $i+1$ (or $N$ and 1, if $i = N$) reaches agreement iff the $i$th row $\psi_i^T$ of the product $\Psi = TQ$ possesses the configuration:
$$\psi_i^T = \begin{bmatrix} \underset{(1)}{*} & \underset{(2)}{0} & \cdots & \underset{(\alpha-1)}{0} & \underset{(\alpha)}{*} & \cdots & \underset{(N)}{*} \end{bmatrix}$$
where $T \in R^{N \times N}$ represents any feasible solution of the matrix equation
$$TL = \Phi = \begin{bmatrix} 1 & -1 & & & \\ & 1 & -1 & & \\ & & \cdots & \cdots & \\ & & & 1 & -1 \\ -1 & 0 & \cdots & \cdots & 1 \end{bmatrix}$$

$Q \in R^{N \times N}$ represents the nonsingular matrix that transforms the Laplacian matrix into the similar Jordan canonical form:
$$Q^{-1}LQ = J = \begin{bmatrix} 0 & & & & \\ & \lambda_2 & * & & \\ & & \lambda_3 & \ddots & \\ & & & \ddots & * \\ & & & & \lambda_N \end{bmatrix}$$
and '$*$' denotes any arbitrary value.

## 3.2. Class 2

Before introducing the condition of Class 2, some serviceable theoretical preparations should be firstly expounded.

*Lemma 3* (Determinant of Block Matrix) [29]: Suppose $A$, $B$, $C$, and $D$ are matrices of dimension $n \times n$, $n \times m$, $m \times n$, and $m \times m$, respectively, then

$$\det\left(\begin{bmatrix} A & 0 \\ C & D \end{bmatrix}\right) = \det(A)\det(D) = \det\left(\begin{bmatrix} A & B \\ 0 & D \end{bmatrix}\right)$$

*Definition 2* (Independent Group): If a subgraph has spanning tree of its own and receives no information, then it is called an *independent group* here.

*Proposition 1:* The spectrum of an independent group of $n$th order coincides with $n$ eigenvalues of the overall graph.

*Proof:* Without loss of generality, suppose that the first $n$ vertices of the graph form an independent group, otherwise the indices can be rearranged. Because the group receives no information, the adjacency matrix can be decomposed into the following triangular configuration:

$$W = \begin{bmatrix} W_{11} & 0 \\ W_{21} & W_{22} \end{bmatrix}$$

where $W_{11}$ represents the subgraph of the independent group and $W_{22}$ the subgraph of the remaining vertices. Correspondingly,

$$L = \begin{bmatrix} L_{11} & 0 \\ L_{21} & L_{22} \end{bmatrix}$$

According to Lemma 3, the eigen-polynomial of $L$ is

$$|\lambda I_N - L| = |\lambda I_n - L_{11}||\lambda I_{N-n} - L_{22}|$$

Thus, the spectrum of the independent group coincides with $n$ eigenvalues of the overall graph. □

*Proposition 2:* The agents associated with an $n$th order independent group of the system achieves agreement if and only if $A - \lambda_i F$ ($i = 1, 2, ..., n$; $\lambda_i \neq 0$) are Hurwitz, where $\lambda_i$ ($i = 1, 2, ..., n$; $\lambda_i \neq 0$) are the eigenvalues of the overall Laplacian matrix of the system that correspond to the independent group. Besides, if an independent group achieves agreement, the trajectories of the agents converge to a solution of the dynamical equation $\dot{\xi} = A\xi$.

*Proof:* According to Proposition 1, the spectrum of the independent group coincides with $n$ eigenvalues of the overall graph; meanwhile, there is a single zero eigenvalue among them since the independent group includes a spanning tree of its own. Based on Lemma 1, it can be implied that the agents associated with the independent group achieve agreement if and only if $A - \lambda_i F$ ($i = 1, 2, ..., n$; $\lambda_i \neq 0$) are Hurwitz.

Without loss of generality, consider the dynamics of agent 1. Since the independent group receives no external information, the difference between $\dot{x}_1$ and $Ax_1$ is

$$\dot{x}_1 - Ax_1 = F \sum_{j=1}^{n} w_{ij}(x_j - x_1)$$

Because the agents 1~n achieve agreement, $\lim_{t \to \infty} \|x_j - x_1\| = 0$ ($j = 1, 2, ..., n$), and as a result, $\lim_{t \to \infty} \left\| F \sum_{j=1}^{n} w_{ij}(x_j - x_1) \right\| = 0$. Consequently, $\lim_{t \to \infty} \dot{x}_1 = Ax_1$. □

*Corollary 1:* If all $A - \lambda_i F$ ($i = 1, 2, ..., N$; $\lambda_i \neq 0$) in a system are Hurwitz, then the agents associated with any independent group achieve agreement.

The condition for Class 2 of non-consensus motions of high-order LTI multi-agent systems is:

1. Some of the elements in

$$\{A - \lambda_i F \mid i = 1, 2, ..., N; \lambda_i \neq 0\}$$

   are not Hurwitz, where $\lambda_i$ ($i = 1, 2, ..., N$; $\lambda_i \neq 0$) are the nonzero eigenvalues of the Laplacian matrix $L(G)$;

2. There is no spanning tree in the graph of system.

In addition, this class can be classified into two types of cases with regard to whether the matrix $A$ is Hurwitz.

### A. $A$ is Hurwitz

*Example 3:* Consider an LTI multi-agent system with the graph having no spanning tree, which is illustrated in Fig. 4.

>>put Fig. 4 about here<<

Fig.4. Graph of system. Default edge weight is 1.

The adjacency matrix is

$$W = \begin{bmatrix} 0 & 0 & 1 & 0 & 1 & 1 \\ 0 & 1 & 0 & 0 & 0 & 1 \\ 0 & 0 & 0 & 0 & 1 & 0 \\ 0 & 1 & 0 & 0 & 0 & 0 \\ 0 & 0 & 0 & 0 & 0 & 0 \\ 0 & 1 & 0 & 0 & 0 & 1 \end{bmatrix}$$

and the Laplacian matrix is

$$L = \begin{bmatrix} 3 & 0 & -1 & 0 & -1 & -1 \\ 0 & 1 & 0 & 0 & 0 & -1 \\ 0 & 0 & 1 & 0 & -1 & 0 \\ 0 & -1 & 0 & 1 & 0 & 0 \\ 0 & 0 & 0 & 0 & 0 & 0 \\ 0 & -1 & 0 & 0 & 0 & 1 \end{bmatrix}$$

with the eigenvalues $\{3, 1, 2, 0, 1, 0\}$.

Suppose that the agents are LTI second-order asymptotically stable systems with

$$A = \begin{bmatrix} -1 & 1 \\ 0 & -2 \end{bmatrix} \text{ and } F = \begin{bmatrix} -0.5 & 0.5 \\ -0.5 & -0.5 \end{bmatrix}$$

For the above LTI multi-agent system, the eigenvalues of state matrix $A$ are $\{-1, -2\}$, and the eigenvalues of $A - \lambda_i F$ $(i = 1, 2, \ldots, 6)$ are

$$\{\pm 0.71i\}$$
$$\{-1.71, -0.30\}$$
$$\{-1, 0\}$$
$$\{-1, -2\} \quad (\lambda_4 = 0)$$
$$\{-1.71, -0.30\}$$
$$\{-1, -2\} \quad (\lambda_6 = 0)$$

respectively. Evidently, some of the elements in $\{A - \lambda_i F \mid i = 1, 2, \ldots, N;\ \lambda_i \neq 0\}$ are not Hurwitz. The state trajectories of Example 3 are shown in Fig. 5.

>>put Fig. 5 about here<<

Fig. 5. State trajectories of Example 3 with $t \in [0, 7]$. Thick dots denote starting positions.

In Fig. 5, agent 3 and agent 5 will gradually converge to a common state trajectory which will converge to the origin with time elapsing. Likewise, agent 4 and agent 6 also will gradually converge to another common state trajectory. This phenomenon is referred to as group clustering. Agent 2 has its own unique state trajectory. Agent 1 will oscillate persistently.

In fact, according to Definition 2, two independent groups can be found in the graph as shown in Fig. 6.

>>put Fig. 6 about here<<

Fig. 6. Two independent groups of the graph.

The Laplacian matrices of independent groups 1 and 2 are:

$$L_1 = \begin{bmatrix} 1 & -1 \\ 0 & 0 \end{bmatrix}, \quad L_2 = \begin{bmatrix} 1 & 0 & -1 \\ -1 & 1 & 0 \\ -1 & 0 & 1 \end{bmatrix}, \text{ respectively.}$$

The eigenvalues for the Laplacian matrix of independent group 1 and independent group 2 are $\{1, 0\}$ and $\{1, 2, 0\}$, respectively. The eigenvalues of $A - \lambda_i F$ in independent group 1 are $\{-1.71, -0.30\}$ and $\{-1, -2\}$. The eigenvalues of $A - \lambda_i F$ in independent group 2 are $\{-1.71, -0.30\}$, $\{-1, 0\}$ and $\{-1, -2\}$. Evidently, all of the elements in $\{A - \lambda_i F | \lambda_i \neq 0\}$ for independent group 1 are Hurwitz, whilst some of the elements in $\{A - \lambda_i F | \lambda_i \neq 0\}$ of independent group 2 are not Hurwitz.

According to the above analysis and Proposition 1, all agents in independent group 1 will achieve agreement, with the common state trajectory converging to the origin as time elapses. In contrast, the phase of motion for independent group 2 by itself belongs to the above-mentioned Class 1. The affiliation of agent 1 is undetermined, depending on the joint attraction from independent group 1 and independent group 2.

### B. $A$ is not Hurwitz

*Example 4:* Consider an LTI multi-agent system with graph which has no spanning tree, illustrated in Fig. 4, with the edge weight between agents 3 and 5 adjusted to 0.3.

Correspondingly the adjacency matrix is now

$$W = \begin{bmatrix} 0 & 0 & 1 & 0 & 1 & 1 \\ 0 & 1 & 0 & 0 & 0 & 1 \\ 0 & 0 & 0 & 0 & 0.3 & 0 \\ 0 & 1 & 0 & 0 & 0 & 0 \\ 0 & 0 & 0 & 0 & 0 & 0 \\ 0 & 1 & 0 & 0 & 0 & 1 \end{bmatrix}$$

and the Laplacian matrix is

$$L = \begin{bmatrix} 3 & 0 & -1 & 0 & -1 & -1 \\ 0 & 1 & 0 & 0 & 0 & -1 \\ 0 & 0 & 0.3 & 0 & -0.3 & 0 \\ 0 & -1 & 0 & 1 & 0 & 0 \\ 0 & 0 & 0 & 0 & 0 & 0 \\ 0 & -1 & 0 & 0 & 0 & 1 \end{bmatrix}$$

with the eigenvalues being $\{3, 1, 2, 0, 0.3, 0\}$.

Suppose that the agents are LTI second-order unstable systems with

$$A = \begin{bmatrix} 1 & 1 \\ -2 & 0 \end{bmatrix} \text{ and } F = \begin{bmatrix} 7 & 5 \\ -4 & -1 \end{bmatrix}$$

For the above LTI multi-agent system, the eigenvalues of state matrix $A$ are $\{0.50 \pm 1.32i\}$, and the eigenvalues of $A - \lambda_i F$ $(i = 1, 2, ..., 6)$ are

$$\{-8.50 \pm 2.78i\}$$
$$\{-4.56, -4.44\}$$
$$\{-7.00, -4.00\}$$
$$\{0.50 \pm 1.32i\} \quad (\lambda_4 = 0)$$
$$\{-1.34, 0.54\}$$
$$\{0.50 \pm 1.32i\} \quad (\lambda_6 = 0)$$

respectively. Evidently, some of the elements in $\{A - \lambda_i F \mid i = 1, 2, ..., N; \lambda_i \neq 0\}$ are not Hurwitz. The state trajectories of Example 4 are shown in Fig. 7.

>>put Fig. 7 about here<<

Fig. 7. State trajectories of Example 4 with $t \in [0, 3.6]$. Thick dots denote starting positions.

In Fig. 7, agents 2, 4 and 6 will gradually aggregate to a common state trajectory which will be away from origin with time elapsing. This phenomenon is referred to as group clustering. Differently, agent 1, agent 3 and agent 5 have their own separate state trajectories.

Similarly, according to Definition 2, two independent groups can be found in the graph shown in Fig. 6.

The Laplacian matrix of independent group 1 and independent group 2 are:

$$L_1 = \begin{bmatrix} 0.3 & -0.3 \\ 0 & 0 \end{bmatrix}, \quad L_2 = \begin{bmatrix} 1 & 0 & -1 \\ -1 & 1 & 0 \\ -1 & 0 & 1 \end{bmatrix}, \text{ respectively.}$$

The eigenvalues for the Laplacian matrix of independent group 1 and independent group 2 are $\{0.3, 0\}$ and $\{1, 2, 0\}$, respectively. The eigenvalues of $A - \lambda_i F$ ($i = 1, 2$) of independent group 1 are $\{-1.34, 0.54\}$ and $\{0.50 \pm 1.32i\}$. The eigenvalues of $A - \lambda_i F$ ($i = 1, 2, 3$) of independent group 2 are $\{-4.56, -4.44\}$, $\{-7.00, -4.00\}$ and $\{0.50 \pm 1.32i\}$. Evidently, some of the elements in $\{A - \lambda_i F | \lambda_i \neq 0\}$ of independent group 1 are not Hurwitz, and all of the elements in $\{A - \lambda_i F | \lambda_i \neq 0\}$ of independent group 2 are Hurwitz.

Based on the above analysis and Proposition 1, the feature of motion of independent group 1 is known to be indefinite and belongs to the above-mentioned Class 1. All agents in independent group 2 will achieve agreement and the common state trajectory will be away from the origin with time elapsing. The affiliation of agent 1 is unknown and depends on the joint attraction from both the independent group 1 and independent group 2.

### 3.3. Class 3

The condition for the third class of non-consensus motions of high-order LTI multi-agent systems is:
1. The state matrix $A$ of each agent is not Hurwitz;
2. All of the elements in
$$\{A - \lambda_i F | i = 1, 2, ..., N; \lambda_i \neq 0\}$$
are Hurwitz, where $\lambda_i$ ($i = 1, 2, ..., N; \lambda_i \neq 0$) are the nonzero eigenvalues of the Laplacian matrix $L(G)$;
3. There is no spanning tree in the graph of system.

*Example 5:* Consider an LTI multi-agent system where the graph topology has no spanning tree, which is illustrated in Fig. 4.

Suppose that all agents are LTI second-order unstable systems with
$$A = \begin{bmatrix} 1 & 5 \\ -0.4 & 0 \end{bmatrix} \text{ and } F = \begin{bmatrix} -1.67 & 1.33 \\ 22.85 & 3.16 \end{bmatrix}$$

For the above LTI multi-agent system, the eigenvalues of state matrix of each agent

$A$ are $\{0.50 \pm 1.32i\}$, and the eigenvalues of $A - \lambda_i F$ $(i = 1, 2, \ldots, 6)$ are

$$\{-1.74 \pm 3.11i\}$$
$$\{-0.25 \pm 8.77i\}$$
$$\{-0.99 \pm 8.91\}$$
$$\{0.50 \pm 1.32i\} \quad (\lambda_4 = 0)$$
$$\{-0.25 \pm 8.77i\}$$
$$\{0.50 \pm 1.32i\} \quad (\lambda_6 = 0)$$

respectively. Evidently, all of the elements in $\{A - \lambda_i F \mid i = 1, 2, \ldots, N; \lambda_i \neq 0\}$ are Hurwitz. The state trajectories of Example 5 are shown in Fig. 8.

>>put Fig. 8 about here<<

Fig. 8. State trajectories of Example 5 with $t \in [0, 5]$. Thick dots denote starting positions.

In Fig.8, agents 3 and 5 will gradually aggregate to a common state trajectory. Likewise, agents 2, 4 and 6 also will gradually aggregate to another common state trajectory. This phenomenon is referred to as group clustering. Differently, agent 1 will has its own separate state trajectory.

Similarly, according to Definition 2, two independent groups can be found in the graph being shown in Fig. 6.

According to Proposition 1, all agents in independent group 1 and independent group 2 will achieve agreement respectively with both the two different common state trajectories diverging from the origin as time elapses. The affiliation of agent 1 is unknown and depends on the joint attraction from independent group 1 and independent group 2 simultaneously.

Under the condition of Class 3, group clustering will be certain to occur.

## 4. Conclusion

The non-consensus problem of high order homogeneous LTI multi-agent systems has been investigated in this paper, mainly based on the necessary and sufficient conditions for consensus achievement. The different cases of non-consensus motions of LTI multi-agent systems can be classified into two categories roughly: group clustering and mutual repulsion. Further, non-consensus motions can be concretely categorized into three classes, each with distinct features. For instance, the dynamical features for Class 1 are complicated and indefinite; for Class 2, the motions of independent groups are relatively simpler to anticipate; the group clustering phenomenon is certain to appear in Class 3. The theoretical conditions and typical numerical examples of each class of motion phases are presented.

Relevant knowledge could enrich the literature by reinforcing our understanding of cause and mechanism for both consensus and non-consensus phenomena. The current research only takes a first step along the direction of revealing the knowledge on various non-consensus problems.

## Acknowledgments

This work is supported by National Natural Science Foundation (NNSF) of China (Grants 61263002 & 61374054), by Fundamental Research Funds for the Central Universities of Northwest University for Nationalities (Grants 31920160003 & 31920170141), and by Program for Young Talents of State Ethnic Affairs Commission (SEAC) of China (Grant 2013-3-21).

# Figures

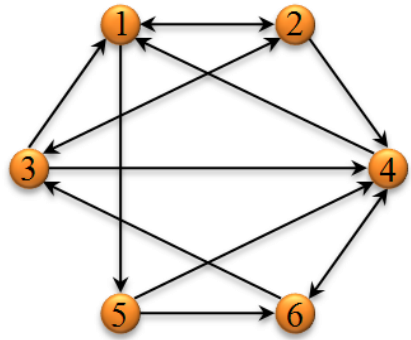

Fig. 1

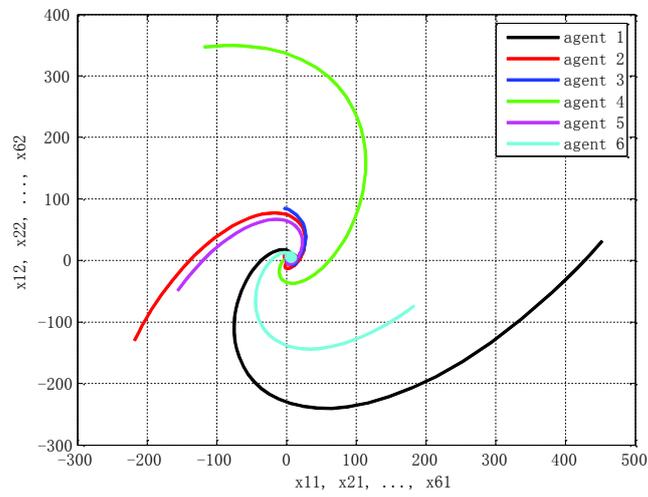

Fig. 2

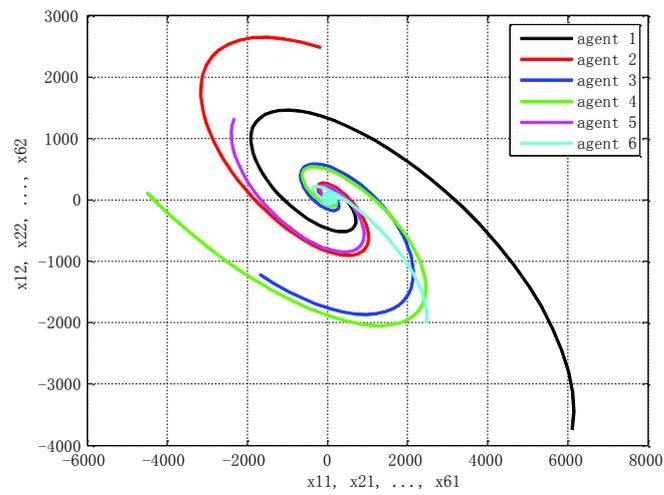

Fig. 3

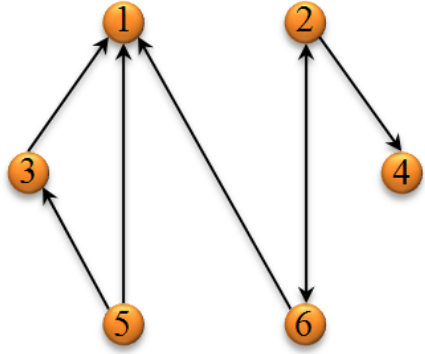

Fig. 4

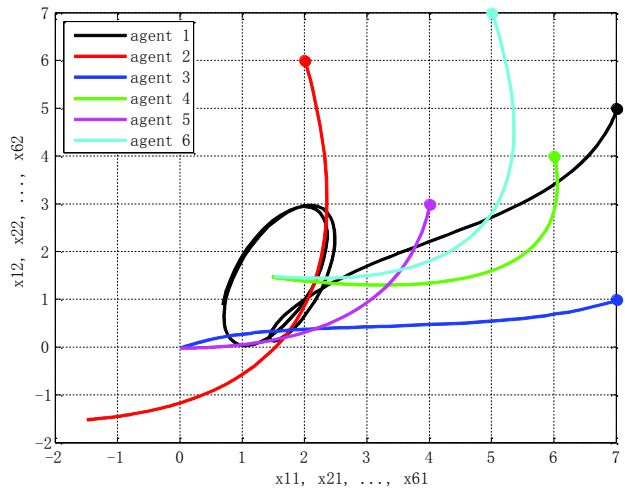

Fig. 5

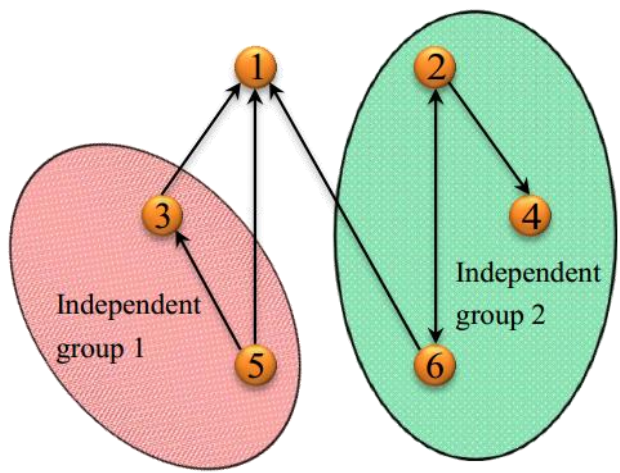

Fig. 6

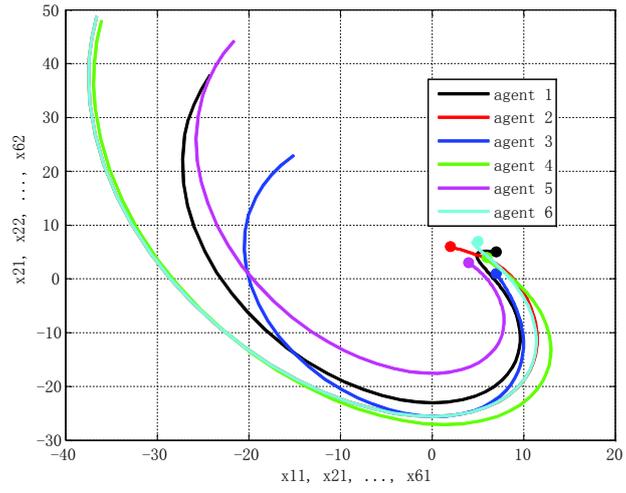

Fig. 7

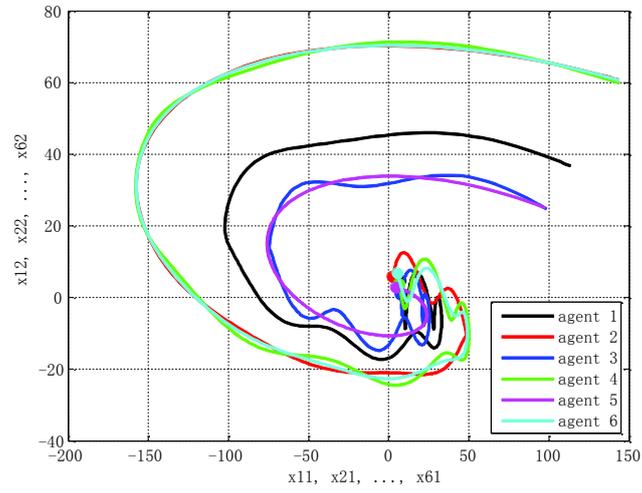

Fig. 8